\documentclass[a4paper]{jpconf}
\usepackage{graphicx}
\usepackage{hyperref}

\usepackage[outdir=./]{epstopdf}

\def\asec{\ifmmode ^{\prime\prime}\else$^{\prime\prime}$\fi}

\def\it{\sl}
\def\degs{\ifmmode ^{\circ}\else$^{\circ}$\fi}
\def\amin{\ifmmode ^{\prime}\else$^{\prime}$\fi}
\def\asec{\ifmmode ^{\prime\prime}\else$^{\prime\prime}$\fi}
\def\fdg{\hbox{$.\!\!^\circ$}}          % Fractions of degrees
\def\farcs{\hbox{$.\!\!^{\prime\prime}$}}  % Fractions of arcseconds
\def\degs{\ifmmode ^{\circ}\else$^{\circ}$\fi}
\def\amin{\ifmmode ^{\prime}\else$^{\prime}$\fi}
\def\farcm{\hbox{$.\mkern-4mu^\prime$}}
\def\eqalign#1{\null\,\vcenter{\openup1\jot \m@th
   \ialign{\strut\hfil$\displaystyle{##}$&$\displaystyle{{}##}$\hfil
   \crcr#1\crcr}}\,}
\def\j0633{J0633}

\def\chan{\textit{Chandra}}

\newcommand{\avk}[1]{{#1}}
\newcommand{\aad}[1]{{#1}}
\usepackage{xcolor}

\begin{document}

\title{\aad{\chan\ measurements of the proper motion of the $\gamma$-ray pulsar J0633+0632 } }
\author{A A Danilenko, A V Karpova and Yu A Shibanov}

\address{Ioffe Institute, Politekhnicheskaya 26, St.~Petersburg, 194021, Russia}

\ead{danila@astro.ioffe.ru}

\begin{abstract}
\aad{We measured the proper motion of a $\gamma$-ray radio-quiet pulsar J0633+0632 using \chan\ observations performed in 2009 and 2017. The measured proper motion is $53\pm15$ mas~yr$^{-1}$. We found that the proper motion direction does not follow the extension of the J0633+0632 pulsar wind nebula. The J0633+0632 pulsar wind nebula therefore can be a jet-like feature or a misaligned outflow. We also discuss a possible birth cite of the pulsar.}
\end{abstract}

\section{Introduction}
\label{s:intro}

The radio-quiet pulsar J0633+0632 (hereafter \j0633) was discovered in $\gamma$-rays by the \textit{Fermi} observatory \cite{abdo2009}. The pulsar has a period $P=297.4$ ms, a characteristic age $\tau = 59.2$ kyr, a spin-down luminosity $\dot{E}= 1.2 \times 10^{35}$ erg~s$^{-1}$, and a surface magnetic field $B=4.9 \times 10^{12}$ G \cite{abdo2013ApJS}. Follow-up \textit{Chandra} observations showed that \j0633 is one of the brightest pulsars in X-rays (its flux $F_X \sim$ $10^{13}$ erg~cm$^{-2}$~s$^{-1}$) among those discovered by \textit{Fermi} \cite{ray2011}. The pulsar X-ray spectrum shows signs of thermal emission from the neutron star surface as well as non-thermal emission of the pulsar magnetosphere origin \cite{ray2011,danilenko2015,karpova2017}.

\begin{figure}
  \begin{center}
    \includegraphics[scale=0.55]{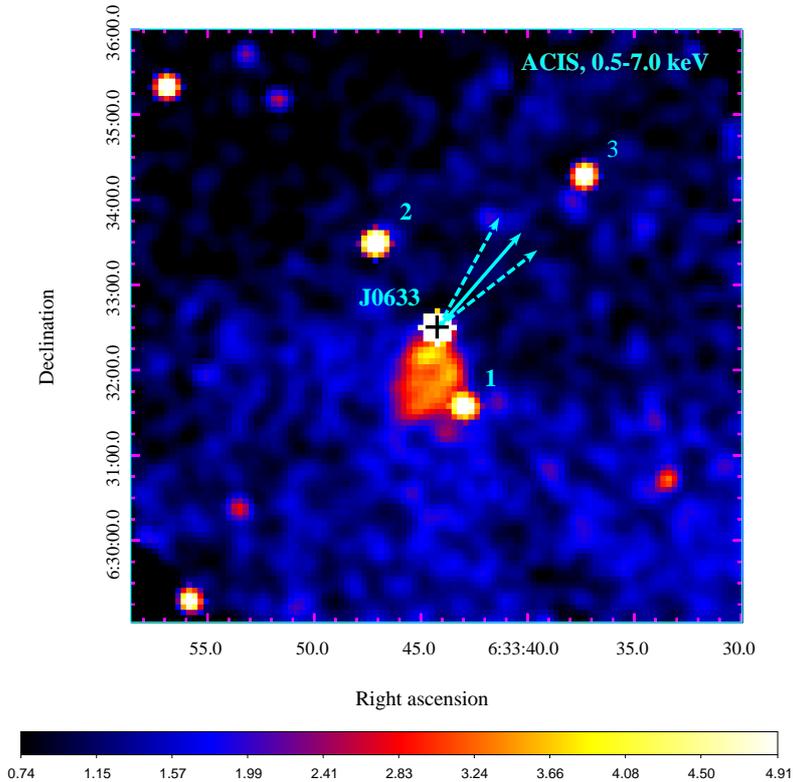}
  \end{center}
  \caption{\avk{7\amin $\times$ 7\amin} exposure-corrected image of the PSR J0633+0632 field combined from all observations listed in table~\ref{t:obs}. The cross marks the pulsar position. Solid and dashed arrows show the direction of the pulsar proper motion and its 1$\sigma$ uncertainties. Numbers mark background stars used for astrometry. The intensity is given in 10$^{-7}$ ph~cm$^{-2}$~s$^{-1}$ pixel$^{-1}$.}
  \label{f:acis_trim}
\end{figure}

\begin{figure}
  \begin{center}
    \includegraphics[scale=0.6]{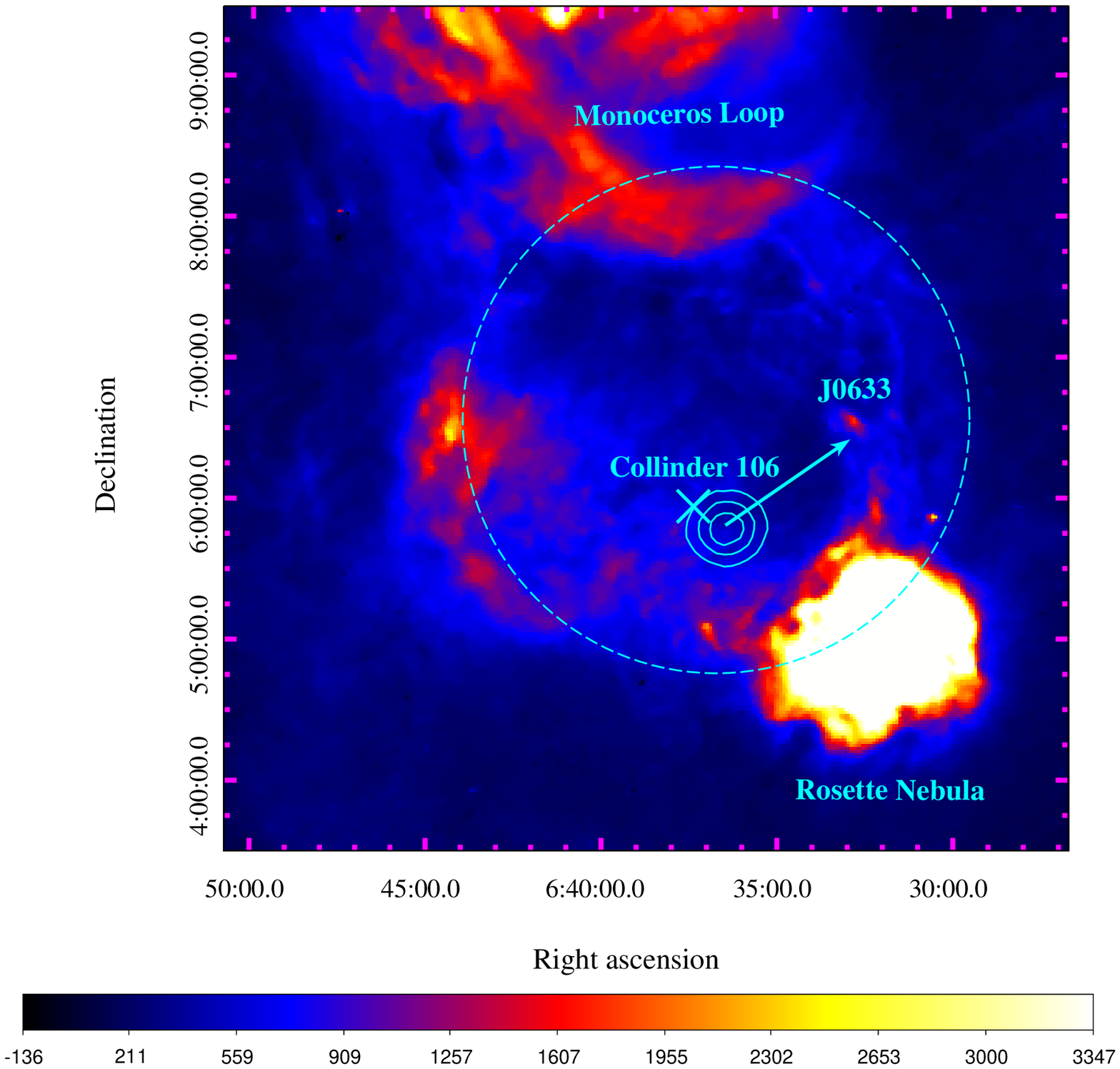}
  \end{center}
  \caption{\avk{6\degs $\times$ 6\degs} H$\alpha$ image of the Monoceros Loop region from the Southern H-Alpha Sky Survey Atlas \cite{gaustad2001}. The arrow shows a trajectory between the pulsar birth-cite and its current position. Contours enclose credible (40, 68, and 95 per cents) regions of the pulsar birth-cite coordinates. The `X' symbol marks the position of the open stellar cluster Collinder 106 and the dashed circle shows an extend of the Monoceros Loop SNR.}
  \label{f:mono}
\end{figure}

There is an elongated pulsar wind nebula (PWN) extended southwards the pulsar (figure~\ref{f:acis_trim}). Similar tail-like PWNe are usually observed behind fast-moving pulsars, as, for instance, the Mouse nebula around PSR J1747$-$2958 \cite{hales2009}. We can assume that \j0633 might as well be moving fast. From a lifetime of electrons emitting synchrotron X-rays ($\approx 10^{3}$ yr) and the \j0633 PWN elongation ($\approx$ 1\farcm3), one can expect the proper motion \avk{$\mu \approx$ 80 mas~yr$^{-1}$ \cite{danilenko2015}.} If the pulsar were moving with such a proper motion, assuming the spin-down age, the angular shift from the pulsar birth-cite would be \avk{$\approx$ 1\fdg3.} Curiously, there is a young and active star-forming region, the Rosette nebula, seen at about the same angular distance from the pulsar (figure~\ref{f:mono}).

To measure the pulsar proper motion and examine the presumed association of the pulsar with the Rosette nebula, we performed new \textit{Chandra} observations of \j0633.

\section{Observations}
\label{s:obs}

We observed \j0633 with \textit{Chandra} on 2017 December 10 \avk{(table~\ref{t:obs})}. Due to some technical reasons, observations were  divided into halves. For reference, we also used the archival \textit{Chandra} data on \j0633 obtained about 8 yr earlier (table~\ref{t:obs}). We reprocessed all data sets using \texttt{CIAO v.4.9} \texttt{chandra\_repro} \avk{tool}. 

In figure~\ref{f:acis_trim}, we present an image of the \j0633 vicinity combined from all data sets obtained  using the \texttt{CIAO merge\_obs} tool. For point-source detection and position measurements, we used the \texttt{CIAO wavdetect} tool \avk{and images in the 0.5--7 keV band.} \aad{For \textit{Chandra}, besides statistical errors returned by \texttt{wavdetect}, there is a systematic positioning inaccuracy of $\approx$~0\farcs07 \cite{motch2009} which we also took into account.} There are luckily at least three background stars seen around \j0633, marked by numbers in figure~\ref{f:acis_trim}, which are detected with signal-to-noise ratios greater than 3 in all three data sets. 

\aad{We then used the measured positions of these three background stars, to find an optimal astrometric solution which aligns all the \textit{Chandra} images together. For the solution, we used a linear transformation which allows for the image shift, rotation and also for an offset between positions of the pulsar at the two epochs.}  

\begin{table}
\caption{\textit{Chandra} observations of J0633+0632.}
\label{t:obs}
%\footnotesize\rm
\centering
\begin{tabular}{cccccc}
\br
Date & Exposure time (s) & Instrument & MJD & Observation ID & PI\\
\mr
2009-12-11 & 19920 & ACIS-S & 55176.8 & 11123 & Roberts\\
2017-12-10 & 14874 & ACIS-S & 58097.3 & 19165 & Danilenko\\
2017-12-10 & 14874 & ACIS-S & 58097.8 & 20876 & Danilenko\\
\br
\end{tabular}
\end{table}

\section{Results and discussion}
\label{s:res}

In table~\ref{t:pm}, we show best-fit values and 1$\sigma$ credible intervals for the offset, an absolute value of the pulsar proper motion, $\mu$, and the proper-motion position angle. Note that the proper motion absolute value is in a reasonable accordance with initial crude estimates. 

We also show the pulsar transverse velocity scaled by a distance, $D_{\rm kpc}$, to the pulsar measured in 1 kpc in table~\ref{t:pm}. The \j0633\ distance, \avk{0.7--2.2} kpc, was estimated using empirical relations between interstellar absorption and distance \cite{karpova2017}. \aad{This distance range corresponds to transverse velocities of 130--730 km~s$^{-1}$ which is in agreement with the empirical distribution of pulsar velocities \cite{hobbs2005}}. 

\begin{table}
\caption{Proper motion of \j0633.}
\label{t:pm}
%\footnotesize\rm
\centering
\begin{tabular}{cccc}
\br
Offset & Position Angle & $\mu$ & $v_{\rm psr}$\\
\mr
\aad{0\farcs42 $\pm$ 0\farcs12} & \aad{$-$48\degs $\pm$ 16\degs} & \aad{53 $\pm$ 15} mas~yr$^{-1}$ & \aad{(260 $\pm$ 70)}$D_{\rm kpc}$ km~s$^{-1}$ \\
\br
\end{tabular}
\end{table}

However, the pulsar proper motion direction is unexpected. In figure~\ref{f:mono}, we present an image of the Rosette Nebula and the Monoceros Loop (G205.5+0.5) supernova remnant (SNR). The latter is a large shell-type middle-aged (30--150 kyr) SNR which may interact with the Rosette Nebula \cite{borka2009,xiao2012}. Using the pulsar proper motion, we constrained coordinates of the pulsar birth cite. Credible regions of the pulsar coordinates 60 kyr back (the characteristic age) are shown in figure~\ref{f:mono}.    

As it is seen from figure~\ref{f:mono}, the \j0633 birth-cite cannot be in the Rosette Nebula. It is also far from the Monoceros Loop center. On the other hand, the credible regions overlap with the open stellar cluster Collinder 106 which can be found in the WEBDA data base. It has an age $t\approx$ 5.5 Myr, a distance $D \approx$ 1.6 kpc and a diameter $d \approx$  1\degs. \aad{The cluster distance is compatible with the \j0633\ distance estimates and it is old enough so it had time to produce a pulsar.}

\avk{The extension of the \j0633\ PWN does not follow the proper motion direction (figure~\ref{f:acis_trim}). In other words, the nebula may have a jet-like/misaligned outflow morphology (see \cite{reynolds2017,kargaltsev2017}). There are a number of examples of such morphology. For instance, PSR B2224+65 powers a long X-ray jet inclined by $\sim$118\degs\ to the pulsar's proper motion direction \cite{hui2007}. Another interesting object is the Lighthouse nebula powered by PSR J1101$-$6101 which shows jet-like structures extended almost perpendicular to the elongation of the bow-shock PWN \cite{pavan2016}. 
%Or, for instance, 
One more example is PSR J1509$-$5850
%X-ray observations of PSR J1509$-$5850 revealed 
which has two `tails'
%one aligned with the presumed pulsar's proper motion direction (which is inferred from the shape of the compact nebula) 
%and the other inclined by $\approx33$\degs\ to it 
with the angle of $\approx150$\degs\ between them \cite{klinger2016}.
These features can be attributed to the synchrotron emission produced 
by high-energy electrons which were accelerated at the PWN termination shock and escaped into the interstellar medium (ISM) with magnetic fields. 
%\cite{bandiera2008,barkov2019}.  
In this case, the mentioned PWN structures  
%Moving along the interstellar medium magnetic field, these electrons 
likely reveal  the ISM magnetic field geometry \cite{bandiera2008,barkov2019}.

Misalignment between the proper motion direction and PWN elongation 
can also be provided by the reverse shock from the surrounding SNR. 
Such explanation was suggested for an extended PWN of PSR B1823$-$13 \cite{pavlov2008}. 
However, we have not yet found an SNR which can cause such misalignment in the case of \j0633.}

\ack{This research has made use of data obtained from the Chandra Data Archive and software provided by the Chandra X-ray Center (CXC).}

\section*{References}
\bibliographystyle{iopart-num}
\bibliography{refj0633}

\end{document}